\documentclass[10pt]{article}
\usepackage{amets,amssymb,amsmath,amstext,lineno,color,xcolor,comment,grffile,array,multirow,makecell,nicefrac}
\usepackage[percent]{overpic}
\usepackage{hyperref,graphicx,multicol,pdfpages}
\hypersetup{colorlinks = true, allcolors = blue, allbordercolors = white}
\usepackage[round,authoryear]{natbib}
\usepackage{float,rotating,epic,tabularx}
\bibpunct[,]{(}{)}{;}{a}{}{,}
\begin{document}
\renewcommand{\baselinestretch}{1.0}\small\normalsize

\begin{center}
\vspace{0.2in}
{\Large On completing a measurement model by symmetry\\}
\vspace{0.3in}

\small
\noindent
Richard E. Danielson\\
Bedford Institute of Oceanography, Fisheries and Oceans Canada,\\
Halifax, Nova Scotia, Canada (rick.danielson@dfo-mpo.gc.ca)\\
\vspace{0.1in}

\normalsize
\vspace{0.05in}
Keywords: model error, equation error, representation error\\
\vspace{0.2in}

\large
Abstract\\
\end{center}
\normalsize

An appeal for symmetry is made to build established notions of
specific representation and specific nonlinearity of measurement
(often called ``model error'') into a canonical linear regression
model.  Additive components are derived from the trivially complete
model $M = m$.  Factor analysis and equation error motivate
corresponding notions of representation and nonlinearity in an
errors-in-variables framework, with a novel interpretation of terms.
It is suggested that a modern interpretation of correlation involves
both linear and nonlinear association.

\vspace{0.15in}
\noindent\rule{\textwidth}{0.5pt}
\vspace{-0.05in}

%\newpage
\begin{multicols}{2}
\normalsize

\section{Introduction}
\label{introducing}

A regression model provides an imprecise framework of inference, but
not just because we happen to employ exploratory measurements, whose
causes are only partially understood.  Notions of measurement error
and model error are both in play.  Notably, \citet{Kruskal_1988} warns
us to be cautious of what we might describe as correlated model error.
Yet how is {\it caution} expressed in a model?  We address the related
question of whether additive models are complete with respect to the
measurements that they are a model of.  One expression that we propose
to include is equation error \citep{Fuller_1987,
  Carroll_Ruppert_1996}, which is the notion that a linear
relationship between measurements is ad hoc, even in the absence of
measurement error.  Prior expressions have taken equation error as a
single identifiable term \citep{Kipnis_etal_1999, Carroll_etal_2006};
here, we take it in parts, as components of two separate terms.

We also consider whether the common distinction between a sample and a
population has an expression in model building.  This is the notion of
specific representation, where we are exploring measurements whose
{\it causal support} is limited, perhaps in part by design.  It seems
tautological that every sample is imperfect, and for every imperfect
sample there is representation error, but this says little about the
difference between two or more imperfect subsets.  Thus, we draw on
factor analysis \citep{Kline_2015} to motivate a more symmetric
treatment of two specific representations.  The resulting expression
is taken as part of a single identifiable term, but we have reason to
interpret equation error and representation error as partially
overlapping.

An appeal is made to explicitly accommodate notions of nonlinearity
and representation in a linear model \citep{Lehmann_2008}.  The
errors-in-variables model \citep{Fuller_2006, Dunn_2011} provides a
canonical framework.  From the outset, symmetry is called on to yield
a division of terms that is specific to our measures, and seems
complete and consistent with an exploration of what we seek to
measure.  The next section describes this model building exercise and
the notions that we ascribe to each term.  Conclusions are given in
the final section, with a comment on distance correlation
\citep{Edelmann_etal_2021} and challenges of an errors-in-variables
solution.
% Conclusions are given in the final section that may restrict our inferences.
% Following \citet{Edelmann_etal_2021}, Section~\ref{solving} discusses

\section{Model building}
\label{building}

We begin with the final form of the model that we wish to derive.
Calibrated ($C$) and uncalibrated ($U$) measurements are on the left
hand side (LHS), and on the right hand side (RHS) are shared truth
($t$) and its additive ($\alpha_U$) and multiplicative ($\beta_U$)
agreement in $U$.  Included are additive shared error ($\epsilon$) and
an unshared residual ($\epsilon_C$ or $\epsilon_U$), as in
\begin{eqnarray}
  \begin{array}{r} C \\ U \end{array}
  \begin{array}{c} = \\ = \end{array}
  \begin{array}{l} \color{white}{\alpha_U + \beta_U} \color{black} t + \epsilon + \epsilon_C \\
                                 \alpha_U + \beta_U                t + \epsilon + \epsilon_U. \end{array}
  \label{eiv}
\end{eqnarray}
The theoretical richness of this model follows from our interpretation
of its terms, but at this point, all we have is a measurement model.
It seems asymmetric to take $U$ to be causal of $C$, following a
traditional approach \citep{Fuller_1987}.  Instead, we make a more
abstract claim that the RHS terms represent causes of both $C$ and
$U$, but as yet, descriptors like ``calibrated'', ``truth'',
``error'', and ``residual'' are undefined.  By ``shared error'', for
instance, we only refer to a RHS duplication of $\epsilon$ (where
``shared'', ``correlated'', and ``associated'' are synonyms), without
an explicit reference to causality, such as that $C$ and $U$ are
independently measured.

Following \citet{Fuller_2006}, the corresponding classic regression
equation is $U = \alpha_U + \beta_U C + \epsilon'$ and the canonical
errors-in-variables model is obtained by writing $\epsilon$ as part of
$\epsilon_C$ and $\epsilon_U$.  We depart further from a traditional
approach by assuming that {\it each term} on the RHS may represent not
just causes that we can explore more easily by $C$ and $U$, but also
those we can't, or don't want to explore in either of them.  Thus, all
variables are simultaneously explanatory and non-explanatory, and a
strict functional (fixed explanatory variable) or structural (random
explanatory variable) distinction is difficult to apply.  However, it
is convenient to think of the explorable (functional) causes as fixed
signal, the more hidden (structural) causes as random noise, and that
a partially different mixture of causes has led to $C$ and $U$.

Because each term on the RHS of a model captures signal and noise, the
trivial model $M = m$ is also a complete model.  By complete, we mean
that under any model partitioning, $m$ remains the union of all
partitioned components.  Assuming that approximate solutions are
identifiable, we opt to divide $M$ and $m$ into well motivated
additive components to build (\ref{eiv}).  We take $C$ and $U$ to be
repeated measurements (two subsets of $M$) that each represent the
same population both nonlinearly and imperfectly, with somewhat
different causal supports.  Otherwise, the division of $M$ is
unspecified.
% except that we presume $C$ and $U$ offer representations that are
% not identical, which implies that the representation of $C$ by $U$,
% or $U$ by $C$, can also be nonlinear and imperfect.
% \footnote{Strictly speaking, a model that is not $M$ itself is
%  difficult to define, but at least numerically, $m$ can be}
% and by ``complete'', we mean that what is commonly called ``model
% error'' is accommodated.  In other words,
% As $m$ remains arbitrarily close to the measurements $M$, there is
% no error in $m$, although $M$ itself includes measurement error.

A model is still complete if an additive division of $m$ is made that
separates the correlated and uncorrelated parts of $M$, as in
\begin{eqnarray}
  \begin{array}{r} C \\ U \end{array}
  \begin{array}{c} = \\ = \end{array}
  \begin{array}{l} s_C + \epsilon_C \\
                   s_U + \epsilon_U. \end{array}
  \label{sepcor}
\end{eqnarray}
Again, signal and noise are both assumed to exist in the shared ($s_C$
and $s_U$) and unshared ($\epsilon_C$ and $\epsilon_U$) terms.  It is
perhaps unusual to accommodate signal in $\epsilon_C$ and
$\epsilon_U$, at least for the classic regression equation
\citep{Lehmann_2008}, but this is an established interpretation in
factor analysis, where it is called the ``specific'' contribution
\citep{Kline_2015}.  Similarly, we claim that $C$ and $U$ have
specific causal supports and representations of the same population.
With {\it different} specific representations, it would be a formal
loss of generality to exclude signal from what is unshared in either
of them.  Moreover, it follows by symmetry that instead of just two or
three components, the most general accommodation is four: shared and
unshared signal and noise.  With a division of measurements $M$ into
the subsets $C$ and $U$, eight components are implied on the RHS of
(\ref{sepcor}), but with signal and noise combined, the model $m$ is
only separated into four terms.

A complete division of $m$ can also be made by separating the linear
and nonlinear parts of $M$, as in
\begin{eqnarray}
  \begin{array}{r} C \\ U \end{array}
  \begin{array}{c} = \\ = \end{array}
  \begin{array}{l} \color{white}{\alpha_U + \beta_U} \color{black} t + n_C \\
                                 \alpha_U + \beta_U                t + n_U. \end{array}
  \label{seplin}
\end{eqnarray}
Again, signal and noise are both assumed to exist in the linear ($t$
and $\alpha_U + \beta_U t$) and nonlinear ($n_C$ and $n_U$) terms.  As
in (\ref{sepcor}), it may be unusual to accommodate signal in $n_C$
and $n_U$, but this is an established notion in the
errors-in-variables literature, where it is usually expressed as a
single term called ``equation error'' \citep{Fuller_1987,
  Carroll_Ruppert_1996, Kipnis_etal_1999, Carroll_etal_2006}.  Here,
we claim that $C$ and $U$ are both nonlinear measures of the same
population.  To say otherwise, and thereby exclude signal from what
may be nonlinear about either of them, would be a formal loss of
generality.  Again, it follows by symmetry that the most general
accommodation is four components: linear and nonlinear signal and
noise.  The RHS of (\ref{seplin}) is thus taken as four terms, each
with signal and noise components.

As a basis for {\it exploring} causality by $C$ and $U$, we interpret
(\ref{eiv}) as a division of $m$ into terms that capture linear
association ($t$ and $\alpha_U + \beta_U t$), nonlinear association
($\epsilon$), and a lack of association ($\epsilon_C$ and
$\epsilon_U$).  As in (\ref{sepcor}) and (\ref{seplin}), this is a
complete division of additive terms if the RHS includes all four
signal and noise components in each of $C$ and $U$; these are the
linear and nonlinear, correlated and uncorrelated components.  By
$\alpha_U$, $\beta_U$, and $t$, the models (\ref{eiv}) and
(\ref{seplin}) provide a generic form of the linear association
between $C$ and $U$.  Because linear association is generic,
$\epsilon$ is both shared and nonlinear by definition.  As in
(\ref{sepcor}), $\epsilon$ is required to provide a generic form of
{\it complete} association.  It follows that the residual terms
($\epsilon_C$ and $\epsilon_U$) capture completely any lack of
association between $C$ and $U$, including any linear and nonlinear
unassociated components that may exist.

When equation error and representation error are aspects of $M$ that
are missing from an incomplete model, then we would expect to describe
them as ``model error'', but in a complete additive model, they are
proper expressions of signal \citep{Kruskal_1988}.  Equation error
typically refers to the nonlinear signal of $\epsilon$ and
$\epsilon_C$, or $\epsilon$ and $\epsilon_U$.  Representation error
typically refers to the signal of $\epsilon_C$ or $\epsilon_U$.  Thus,
it seems difficult to express equation error as a single term, or to
distinguish between equation error and representation error.

\section{Conclusions}
\label{concluding}

An appeal for symmetry is made in deriving and interpreting a complete
additive measurement model.  A complete model is one that includes all
components under any partitioning, even if this partitioning is only
notional.  For example, there is no explicit partitioning of signal
and noise, which are expected to be somewhat different in two set of
measures.  Other partitionings are motivated by ``model error'', or an
incomplete accommodation of the notions of equation error
\citep{Carroll_Ruppert_1996} and representation error.  One novel
aspect is that when complementary notions of specific nonlinearity and
specific representation \citep{Kline_2015} are taken to define
measurements from the outset, a symmetric model formulation highlights
signal components that are not easy to recognize and accommodate as
signal.

This exercise provides a useful characterization of descriptors like
``calibrated'', ``truth'', and ``error''.  If truth and signal are
taken as synonyms, then it seems more appropriate to describe a
variable like $t$ as linear than as truth.  If error and noise are
synonyms, then (\ref{eiv}) can be called both an error model and a
signal model.  Regarding the identity of $C$ as a calibration
reference, this is formally undefined, {\it even if a familiar set of
  measurements is available}.  Strictly speaking, calibration only
applies to corresponding components of association (e.g., linear
calibration only applies to the linear association terms), and not to
all signal components.  The notion that there is signal in every
component of (\ref{eiv}) follows from specific representation, which
is what \citet{Haraway_1988} describes as ``situated knowledges''.

Specific nonlinearity of $C$ and $U$ is another novel aspect, whose
consequence is that variance, covariance, and correlation all include
(by way of $\epsilon$) a nonlinear signal component, which is what
\citet{Pearson_1902} describes as ``personal''.  Specific nonlinearity
is irrelevant without a division of $M$, but our interpretation of
(\ref{eiv}) permits nonlinear association between any two measurement
subsets.  Of course, this is not inconsistent with the practice of
using an appropriate measurement unit, if one is available.
\citet{Mahalanobis_1940} questions the unit, and its function in
allowing us to {\it avoid} systematic differences in measurement.  We
often expect our choice of unit to do just that.  Even if we choose a
unit so as not to contribute to nonlinear association, however, a
contribution could still arise out of differences in the remaining
causal supports of $C$ and $U$.  The use of an appropriate unit might
be expected to minimize differences in causal support, but not
necessarily eliminate them.
% In turn, it is of interest to explore what this nonlinear signal represents.
% Even if measurements are taken in an assigned (linear) unit.  There
% may be many simultaneous causes of $M$ that are active at different
% scales.  Only some causes or scales may be of interest, but a single
% choice of unit, when applied across all scales, may be unable to

Finally, we acknowledge a longstanding need to identify the magnitude
of $\epsilon$ (by any interpretation).  Conventional regression model
solutions often neglect this term using what \citet{Kruskal_1988} and
\citet{Lehmann_2008} describe as the ``assumption of independence''.
Following \citet{Edelmann_etal_2021}, however, a modern definition of
dependence does not place any constraint on correlation.  In practice,
this means that $\epsilon$ cannot be neglected on the basis that
independence seems natural or obvious.  The presence of $\epsilon$
might imply a solution of (\ref{eiv}) that is more ambitious than for
the errors-in-variables model \citep{Fuller_2006, Dunn_2011}, but this
is not necessarily the case \citep{Polya_1973}.  There is the prospect
that if $M$ includes more than two repeated samples ($C$ and $U$), a
numerical solution of (\ref{eiv}) is possible using a complete model
of the longitudinal samples near $C$ and $U$.
\end{multicols}

\bibliographystyle{amets}
\bibliography{refer}

\begin{thebibliography}{14}
\expandafter\ifx\csname natexlab\endcsname\relax\def\natexlab#1{#1}\fi

\bibitem[Carroll and Ruppert(1996)]{Carroll_Ruppert_1996}
Carroll, R.~J., and D.~Ruppert, 1996: The use and misuse of orthogonal
  regression in linear errors-in-variables models. {\em The American
  Statistician\/}, {\bf 50}, 1--6.

\bibitem[Carroll et~al.(2006)Carroll, Ruppert, Stefanski, and
  Crainiceanu]{Carroll_etal_2006}
Carroll, R.~J., D.~Ruppert, L.~A. Stefanski, and C.~M. Crainiceanu, 2006: {\em
  Measurement Error in Nonlinear Models, A Modern Perspective, second
  edition\/}. Chapman and Hall/CRC Press, Boca Raton, Florida, 455~pp.

\bibitem[Dunn(2011)]{Dunn_2011}
Dunn, G., 2011: \textnormal{Method Comparison Studies, {{\it Int. Encyclopedia
  of Statistical Science}}, {M.~Lovric, Ed., Springer}, 815--816,
  doi:10.1007/978-3-642-04898-2\_36}.

\bibitem[Edelmann et~al.(2021)Edelmann, M{\'o}ri, and
  Sz{\'e}kely]{Edelmann_etal_2021}
Edelmann, D., T.~F. M{\'o}ri, and G.~J. Sz{\'e}kely, 2021: On relationships
  between the {Pearson} and the distance correlation coefficients. {\em Stat.
  Prob. Lett.\/}, {\bf 169}, 1--6, doi:10.1016/j.spl.2020.108960.

\bibitem[Fuller(1987)]{Fuller_1987}
Fuller, W.~A., 1987: {\em Measurement Error Models\/}. Wiley, New York, 445,
  doi:10.1002/9780470316665~pp.

\bibitem[Fuller(2006)]{Fuller_2006}
Fuller, W.~A., 2006: \textnormal{{Errors in variables. {\it Encyclopedia of
  Statistical Sciences}, S.~Kotz, C.~B.~Read, N.~Balakrishnan, B.~Vidakovic and
  N.~L.~Johnson, Eds., doi:10.1002/0471667196.ess1036.pub2}}.

\bibitem[Haraway(1988)]{Haraway_1988}
Haraway, D., 1988: Situated knowledges: The science question in feminism and
  the privilege of partial perspective. {\em Feminist Studies\/}, {\bf 14},
  575--599.

\bibitem[Kipnis et~al.(1999)Kipnis, Carroll, Freedman, and
  Li]{Kipnis_etal_1999}
Kipnis, V., R.~J. Carroll, L.~S. Freedman, and L.~Li, 1999: A new dietary
  measurement error model and its application to the estimation of relative
  risk: Application to four validation studies. {\em Am. J. Epidemiol.\/}, {\bf
  150}, 642--651, doi:10.1093/oxfordjournals.aje.a010063.

\bibitem[Kline(2015)]{Kline_2015}
Kline, R.~B., 2015: {\em {Principles and Practice of Structural Equation
  Modeling, Fourth Edition}\/}. Guilford Press, New York, 533~pp.

\bibitem[Kruskal(1988)]{Kruskal_1988}
Kruskal, W., 1988: Miracles and statistics: The casual assumption of
  independence. {\em J. Amer. Statist. Assoc.\/}, {\bf 83}, 929--940.

\bibitem[Lehmann(2008)]{Lehmann_2008}
Lehmann, E.~L., 2008: \textnormal{On the history and use of some standard
  statistical models, {{\it Probability and Statistics: Essays in Honor of
  David A. Freedman}}, {Institute of Mathematical Statistics, Beachwood, Ohio,
  USA}, 114--126, doi:10.1214/193940307000000419}.

\bibitem[Mahalanobis(1940)]{Mahalanobis_1940}
Mahalanobis, P.~C., 1940: Errors of observation in physical measurements. {\em
  {Science and Culture}\/}, {\bf 5}, 443--445.

\bibitem[Pearson(1902)]{Pearson_1902}
Pearson, K., 1902: On the mathematical theory of errors of judgement, with
  special reference to the personal equation. {\em Phil. Trans. Roy. Soc.
  London\/}, {\bf 198}, 235--299.

\bibitem[P\'{o}lya(1973)]{Polya_1973}
P\'{o}lya, G., 1973: {\em How to Solve It: A New Aspect of Mathematical
  Method\/}. Princeton University Press, Princeton, New Jersey, 253~pp.

\end{thebibliography}
\end{document}